\documentclass[conference]{IEEEtran}
\usepackage{cite,amsmath,amssymb,graphicx,booktabs,array}
\usepackage[hidelinks]{hyperref}
\usepackage{dblfloatfix,placeins,balance}
\usepackage[caption=false,font=footnotesize]{subfig}
\graphicspath{{figures/}}
\begin{document}
\title{Human Mobility Sensing Using Wall-Reflected Beamformed Millimeter-Wave Communication Signals}
\author{\IEEEauthorblockN{Tony DiMichele, Raees Matheen Mullah, Christian Lopez Garcia, Zahra Najafi, and Mohammed E. Eltayeb}
\IEEEauthorblockA{Department of Electrical and Electronic Engineering\\
California State University, Sacramento, Sacramento, CA, USA\\
Emails: \{tonydimichele, rmullah, christianlopezgarci, zahra.najafi, mohammed.eltayeb\}@csus.edu}}
\maketitle

\begin{abstract}
Existing millimeter-wave (mmWave) communication
beam patterns spatially weight human-scattered
signals and shape the received-power signatures
induced by human motion. We exploit these
beam-shaped signatures for device-free
walking-pace and walk-direction classification
using a wall-reflected 60.8-GHz link with
closely spaced phased arrays and fixed
communication beamforming vectors.
Human movement alters the interference
between wall-reflected and scattered signals,
producing power oscillations surrounding
a central blockage fade. A beam-dependent
interference model provides a physical
interpretation of these signatures.
Using received-power statistics from a
continuous-tone experiment, we extract
fade duration for walking-pace classification
and pre/post-fade asymmetries for
walk-direction classification.
Measurements comprise 144 crossings from
three participants along perpendicular
and diagonal trajectories.
Under leave-one-participant-out validation,
logistic regression achieves 86.1\% binary
and 79.2\% three-class walking-pace accuracy,
and 77.1\% travel-direction accuracy.
The results demonstrate mobility sensing
using fixed communication beams without
channel estimation, Doppler processing,
or dedicated radar hardware.
\end{abstract}


\section{Introduction}

Millimeter-wave (mmWave) communication
systems rely on directional beamforming
to overcome propagation loss but remain
sensitive to human blockage and scattering
~\cite{rappaport2013,heath2016}.
Although these effects are typically
viewed as communication impairments,
the resulting received-power variations
can provide information about human
movement and create opportunities for
integrated sensing and communications
(ISAC)~\cite{liu2022isac, kumari2018joint}.  In phased-array systems, communication
beamforming codebooks determine the spatial
response of transmitted and received
signals \cite{eltayeb2015opportunistic}. Their main lobes and sidelobes
weight propagation components from
different directions. As a person moves,
human-scattered signals change relative
to the dominant path. The selected beam
patterns shape their interference and
impose characteristic temporal signatures
on the received power. These signatures
can be opportunistically exploited for mobility sensing
without relying on sensing-specific waveform design.

Wireless human sensing at sub-6-GHz
frequencies has been demonstrated using
channel and Doppler measurements for
gesture and activity recognition
~\cite{pu2013wisee,wang2017carm},
walking-direction estimation
~\cite{wu2016widir}, speed estimation
~\cite{zhang2018wispeed}, and tracking
~\cite{qian2018widar2}. At 60~GHz,
human sensing has been demonstrated
using channel impulse responses and
micro-Doppler signatures
~\cite{pegoraro2024}. Beam-dependent
waveform characteristics have also
been exploited for communication-beam
identification~\cite{polese2021deepbeam}.

Other mmWave studies have characterized
human blockage and interference-induced
power oscillations
~\cite{maccartney2016,bhardwaj2021}.
Human-motion sensing at 60 and 300~GHz
has also been demonstrated using wideband
channel sounding and delay-resolved
propagation measurements
~\cite{doeker2025}.
However, such approaches require
channel impulse response acquisition
and extraction of individual
propagation components.
In contrast, our approach exploits
beam-shaped power signatures in a
wall-reflected communication link.
Walking pace and direction are
classified using low-rate received-power
statistics without channel estimation
or multipath-component extraction.

In this paper, we investigate
walking-pace and travel-direction
classification using a wall-reflected
60.8-GHz link with closely spaced transmitter (TX)
and receiver (RX) phased arrays. Fixed
communication beamforming vectors
establish the reflected TX--wall--RX
path, while human movement produces
beam-shaped power signatures.
Our approach exploits these signatures
without channel estimation or
Doppler processing. The main contributions are:
\begin{itemize}

\item We exploit motion-induced power
signatures shaped by fixed communication
beams for  walking-pace
and direction sensing in wall-reflected
mmWave links.

\item We formulate a beam-dependent
interference model and develop
compact sensing features based on fade
duration and pre/post-fade asymmetries.

\item We evaluate the approach across
perpendicular and diagonal walking
trajectories and compare compact and
extended waveform feature representations.

\end{itemize}

\section{Motivation and Sensing Principle}
\label{sec:principle}

Human blockage of mmWave links produces
attenuation and received-power fluctuations
~\cite{maccartney2016,bhardwaj2021}.
Before and after blockage, human-scattered
signals can interfere with the dominant
propagation path and produce power
oscillations~\cite{bhardwaj2021}.
In phased-array systems, communication
beam patterns spatially weight these
components through their main lobes and
sidelobes. As a person moves, changes
in the scattered signals produce
characteristic temporal power signatures
under the selected beamforming vectors.

\begin{figure}[!t]
\centering
\includegraphics[width=.75\columnwidth]
{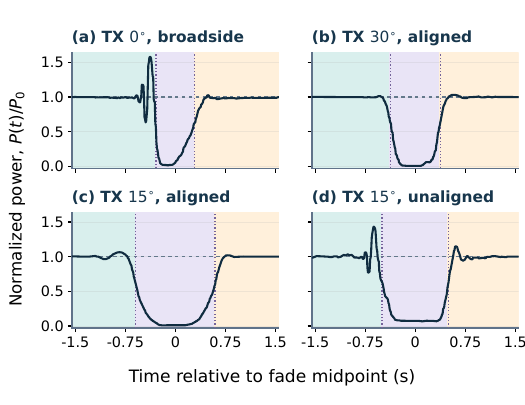}
\caption{Measured LoS hand-blockage
signatures under different TX beam-steering
and pointing configurations. Angles denote
TX steering relative to broadside.
Power is normalized to the unblocked level.
Shaded regions identify pre-blockage (green),
blockage (purple),  and post-blockage (yellow) intervals.}
\label{fig:beam}
\end{figure}

Fig.~\ref{fig:beam} illustrates the
sensitivity of blockage signatures to 60 GHz phased-antenna configuration using a short-range
line-of-sight (LoS) experiment. A hand
crosses the propagation path under
different TX beam-steering vectors and physical
pointing configurations.  Pronounced pre- and post-blockage
oscillations occur in the broadside
and unaligned configurations, whereas
aligned configurations exhibit smoother
transitions.  These measurements show that the
received-power signatures depend on
the selected beamforming codeword
and antenna alignment. The observed
differences suggest that antenna
configuration affects the interference
between human-scattered signals and
the dominant propagation path. The observations therefore motivate exploiting
existing communication beams for mobility
sensing without dedicated sensing beams.
More generally, communication beamforming
codebooks could be designed to enhance
motion-induced power signatures through
controlled sidelobe patterns while
preserving the main communication beam.

\subsection{Wall-Reflected Propagation Model}

The two-ray model in~\cite{bhardwaj2021}
describes interference between direct
and human-reflected signals in LoS links.
We consider a different geometry in
which closely spaced TX and RX phased
arrays communicate through a dominant
wall-reflected path. The propagation
link follows TX--wall--RX rather than
direct TX--RX propagation.
Human movement can perturb either
segment of this reflected path and
introduce additional scattering
interactions involving the body
and the wall.

For a fixed TX/RX beam pair $b$,
the effective channel outside the
central blockage interval is
approximated as 
$
h_b(t)\simeq
\alpha_{w,b}+\alpha_{h,b}(t),
$
where $\alpha_{w,b}$ denotes the
dominant wall-reflected component
and $\alpha_{h,b}(t)$ represents
the time-varying human-scattered
contributions, including interactions
with the wall \cite{bhardwaj2021}. We neglect weaker
background multipath in this
two-component approximation.  Under a dominant specular-reflection
assumption, the wall-reflected
coefficient is approximated as
$
\alpha_{w,b}=
\frac{\lambda\Gamma_w(\theta_w)G_{w,b}}
{4\pi d_w}e^{-jkd_w},
$
where $k=2\pi/\lambda$, $d_w$ is
the TX--wall--RX path length,
$\Gamma_w(\theta_w)$ is the wall
reflection coefficient, and
$G_{w,b}$ is the combined complex
TX/RX antenna response along
the reflected path. The beamforming vectors also determine
the angular weighting of the
human-scattered components.

Consequently, the wall-reflected and
human-scattered components experience
different beam-dependent responses,
which affect their relative amplitudes
and phases at the receiver.  Normalizing the received power by the
dominant wall-reflected power gives
\begin{equation}
\begin{aligned}
\widetilde{p}_b(t)
&=\frac{|\alpha_{w,b}+\alpha_{h,b}(t)|^2}
{|\alpha_{w,b}|^2}
=1+\rho_b^2(t)
+2\rho_b(t)\cos\bigl(\Delta\phi_b(t)\bigr),
\end{aligned}
\label{eq:power}
\end{equation}
where $ \rho_b(t)
=\frac{|\alpha_{h,b}(t)|}
{|\alpha_{w,b}|}, 
\Delta\phi_b(t)
=\arg\{\alpha_{h,b}(t)\}
-\arg\{\alpha_{w,b}\}, $ and $\arg\{\cdot\}$ denotes the phase
of a complex channel coefficient.  The received power therefore depends on the
relative amplitude and phase of the
wall-reflected and human-scattered
components. As a person moves,
the scattered components change,
producing constructive and destructive
interference with the wall reflection.
The selected communication beams
shape these oscillations through
their angular responses. The beam dependence provides the
physical basis for the proposed
sensing approach. Different beam
patterns can produce different power
signatures under similar propagation
conditions because they weight the
scattering paths differently.
For a fixed beam pair, human motion
changes these components without
requiring beam switching.
Thus, existing communication beams
can provide spatially weighted
motion signatures that may be
exploited for sensing.

\begin{figure}[!t]
\centering
\includegraphics[width=.6079\columnwidth]
{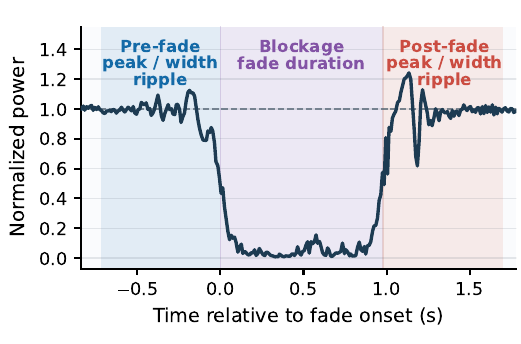}
\caption{Measured wall-reflected
received-power signature at 60.8 GHz
for a right-to-left human crossing.}
\label{fig:event}
\end{figure}

The model describes interference outside
the central blockage interval. During
blockage, obstruction of the dominant
wall-reflected path produces a central
power fade. Fig.~\ref{fig:event}
illustrates a measured crossing with
a central fade surrounded by
interference-related power oscillations.  The central fade reflects the duration
of human obstruction. Faster walking
generally produces shorter fades,
although the duration also depends
on the crossing trajectory and
effective illuminated region.
We therefore use fade duration for
walking-pace classification rather
than absolute velocity estimation. The surrounding power oscillations
provide complementary information
about walking direction. Opposite
walking directions can alter the
sequence of scattering interactions
along the TX--wall and wall--RX paths,
producing asymmetric pre- and post-fade
responses. Accordingly, we investigate
fade duration for walking pace and
pre/post-fade asymmetries for walking
direction. The corresponding features
are defined in
Section~\ref{sec:classification}.

\section{Experimental Platform and Dataset}
\label{sec:experiments}

The experimental platform uses Sivers EVK06002
transmit and receive phased-array
transceivers operating at 60.8~GHz.
A Xilinx KC705 Kintex-7 FPGA interfaces
with ALINX FL9781 digital-to-analog
converter (DAC) and FL9627 analog-to-digital
converter (ADC) boards. The FPGA generates
a continuous 12.5-MHz baseband tone,
which is upconverted to the RF carrier.
At the receiver, the downconverted
in-phase (I) and quadrature (Q) signals
are sampled using a 12-bit ADC operating
at 125~MSPS. The FPGA computes signal
statistics, which are recorded by the
host at approximately 600 reports per second.  The received-power statistic is defined as
\begin{equation}
P(t)=V_{I,\mathrm{rms}}^2(t)
    +V_{Q,\mathrm{rms}}^2(t),
\label{eq:measured_power}
\end{equation}
where $V_{I,\mathrm{rms}}$ and
$V_{Q,\mathrm{rms}}$ denote the root-mean-square (RMS)
voltages of the received I/Q components.
This quantity represents a baseband
power proxy rather than calibrated RF
received power. Each crossing waveform
is normalized as $p(t)=P(t)/P_0$,
where $P_0$ is the median power in
available reference windows located
0.2--1.2~s before and after the crossing.
The reported statistics are grouped
into 10-ms intervals using host timestamps,
and the median within each occupied
interval is retained. Thus, the sensing
method operates on low-rate power
statistics without requiring raw I/Q
waveforms, channel estimation, or
Doppler processing.  The present experiments use a continuous complex
test tone rather than data-bearing traffic.
In an ISAC implementation, similar power
signatures could be acquired using known
reference transmissions during communication
idle intervals or selected OFDM pilot
resources. The latter requires receiver-side
isolation of the pilot component before
power estimation. Both implementations
remain to be evaluated with operational
communication traffic.

\begin{figure}[!t]
\centering
\includegraphics[width=.75\columnwidth]
{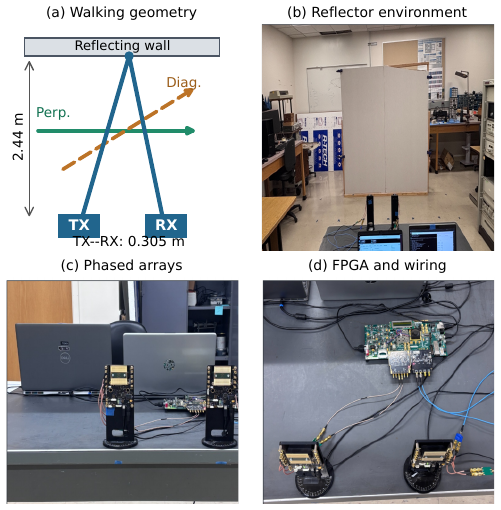}
\caption{Experimental platform and
measurement geometry: (a) wall-reflected
propagation path and perpendicular and
diagonal trajectories, (b) measurement
environment, (c) 60.8-GHz phased-array
transceivers, and (d) FPGA, DAC/ADC,
and RF connections.}

\label{fig:photos}
\end{figure}

Fig.~\ref{fig:photos} shows the
experimental platform and measurement
geometry. The TX and RX arrays are
separated by 0.305~m and positioned
approximately 2.44~m from a reflecting
wall. The nominal walking path is
located 1.83~m from the arrays.
Both phased arrays are directed toward
the wall to establish a dominant
TX--wall--RX propagation path.
Unlike conventional LoS sensing with
separated TX and RX units, this
configuration places both arrays
on the same side of the walking region.

Both arrays use directional beamforming
vectors selected from their communication
codebooks. This configuration is denoted
DD, indicating directional operation
at both TX and RX. The selected vectors
remain fixed throughout the walking
experiments. Their antenna patterns
spatially weight the wall-reflected
and human-scattered propagation components.
Human movement changes the scattered
signals within this fixed spatial response
and produces the beam-shaped power
signatures described in
Section~\ref{sec:principle}.
No beam sweeping or adaptive beam
tracking is performed during a crossing.

\begin{figure}[!t]
\centering
\includegraphics[width=.92\columnwidth]
{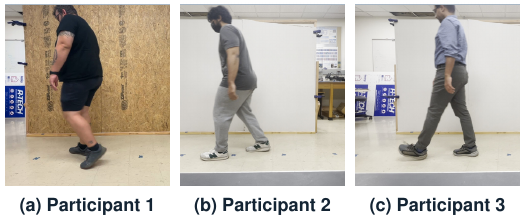}\vspace{-5mm}
\caption{Representative video frames
of participants P1--P3 during the
wall-reflected 60.8-GHz walking
experiments. Video recordings support
crossing-event identification and
travel-direction verification.}
\label{fig:participants}
\end{figure}

Three participants performed instructed
slow, comfortable, and fast walking
trials along perpendicular and diagonal
trajectories. Crossings were recorded
in both left-to-right and right-to-left
directions.  Video recordings
supported crossing-event identification,
and direction labels were verified
against the experimental records.
Representative participant frames
are shown in Fig.~\ref{fig:participants}.

The dataset comprises 144 DD crossings:
56 perpendicular and 88 diagonal.
The perpendicular dataset contains
18 slow, 18 comfortable, and 20 fast
crossings, while the diagonal dataset
contains 24 slow, 30 comfortable, and
34 fast crossings. The combined dataset
therefore includes 42 slow, 48 comfortable,
and 54 fast crossings. Travel directions
are balanced within each geometry, with
28 crossings per direction for
perpendicular walking and 44 per
direction for diagonal walking.  Of the 56 perpendicular crossings,
40 were recorded with a drywall
reflector and 16 with plywood.
These recordings were acquired in
separate sessions and do not constitute
a controlled comparison of reflector
materials. The diagonal measurements
extend the evaluation to a second
walking geometry.  Walking-pace labels correspond to
instructed categories rather than
calibrated velocities. Consequently,
the experiments evaluate walking-pace
and travel-direction classification,
not absolute speed estimation.
The feature definitions and validation
procedures are presented in
Section~\ref{sec:classification}.

\section{Classification and Validation}
\label{sec:classification}

We investigate walking-pace and travel-direction
classification using low-dimensional features
extracted from the received-power signatures.
Each crossing represents one classification
sample. Walking pace is evaluated as binary
(fast versus nonfast) and three-class
(slow, comfortable, and fast) classification,
while travel direction is classified
as left-to-right or right-to-left.  For walking-pace classification, we use
central fade duration $W_{60}$, defined
as the duration of the main contiguous
interval during which normalized received
power falls below 0.60. This feature
characterizes the effective duration
of human obstruction along the
wall-reflected propagation path.
Faster walking generally produces shorter
fades for a fixed geometry. However,
fade duration also depends on the
crossing trajectory, beam patterns,
and effective illuminated region.
We therefore classify instructed
walking paces rather than estimate
absolute velocity.

Travel-direction classification exploits
the asymmetry of power fluctuations
surrounding the central fade. The fixed
communication beams spatially weight
the human-scattered components, while
human movement changes their relative
amplitudes and phases. Opposite walking
directions can alter the sequence of
scattering interactions and produce
different pre- and post-fade responses.
We characterize these signatures using
the two-dimensional feature vector
\begin{equation}
\mathbf{z}_{\mathrm{dir}}=
[A_R,A_S]^{\mathsf T},
\label{eq:features}
\end{equation}
where $A_R$ and $A_S$ represent
ripple-amplitude and transition-slope
asymmetries, respectively.  
Ripple strength is estimated within
1-s intervals immediately before and
after the central fade. A local power
trend is obtained using Gaussian smoothing
with an 80-ms standard deviation.
Let $R_{\mathrm{pre}}$ and $R_{\mathrm{post}}$
denote the RMS amplitudes of the
detrended pre- and post-fade power
fluctuations, respectively.
Ripple-amplitude asymmetry is defined as
\begin{equation}
A_R=
\frac{R_{\mathrm{pre}}-R_{\mathrm{post}}}
{R_{\mathrm{pre}}+R_{\mathrm{post}}}.
\label{eq:ripple_asymmetry}
\end{equation}

The second feature characterizes
the transitions into and out of
the central fade. The falling and
rising slopes, $S_{\mathrm{pre}}$
and $S_{\mathrm{post}}$, are estimated
from changes in local mean normalized
power across intervals separated
by 90~ms. When these measurements
are incomplete, a linear fit to
the available samples within the
transition window is used.
The transition-slope asymmetry is
\begin{equation}
A_S=
\frac{|S_{\mathrm{pre}}|-|S_{\mathrm{post}}|}
{|S_{\mathrm{pre}}|+|S_{\mathrm{post}}|}.
\label{eq:slope_asymmetry}
\end{equation}
Both asymmetry features lie between
$-1$ and $1$ when their denominators
are nonzero.  Positive values indicate
larger pre-fade ripple or slope magnitudes,
while negative values indicate larger
post-fade magnitudes.  Undefined ratios
are treated as missing features.

\begin{figure}[!t]
\centering
\includegraphics[width=.7895\columnwidth]
{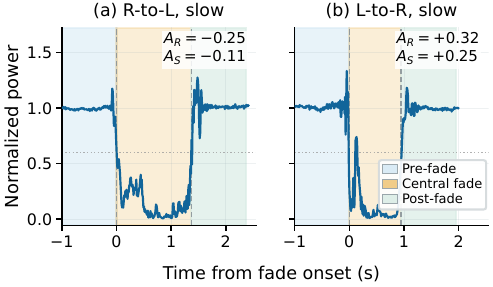}
\caption{Measured normalized-power
signatures for consecutive slow 
crossings in opposite directions by
the same participant using a drywall
reflector. Shaded regions identify
pre-fade, central-fade, and post-fade
intervals. Annotations show ripple-amplitude
asymmetry $A_R$ and transition-slope
asymmetry $A_S$.}
\label{fig:dir_feature_examples}
\end{figure}

Fig.~\ref{fig:dir_feature_examples}
shows measured power signatures from
consecutive slow crossings in opposite
directions by the same participant.
The post-fade ripple and slope magnitudes
are larger in Fig.~\ref{fig:dir_feature_examples}(a),
whereas the pre-fade magnitudes are
larger in (b). These differences produce
opposite signs of $A_R$ and $A_S$,
illustrating the asymmetries exploited
for walking-direction classification.

We use $L_2$-regularized logistic
regression with inverse regularization
parameter $C=1$ as the primary classifier.
Walking-pace classification uses
$W_{60}$, while walking-direction
classification uses
$\mathbf{z}_{\mathrm{dir}}=
[A_R,A_S]^{\mathsf T}$.
The direction classifier learns a
linear decision boundary in the
standardized two-dimensional feature space.  For walking-direction classification,
we also evaluate six feature sets with
dimensions of 1, 2, 3, 5, 14, and 22.
The single-feature model uses ripple
asymmetry $A_R$, while the two-feature
model adds slope asymmetry $A_S$.
The three-feature vector includes
the normalized timing of the fade
minimum. The five-feature vector
further adds in-fade ripple strength
and ripple-active duration asymmetry.  Separately, the 14-dimensional exterior
morphology vector includes peak width,
height, prominence, temporal offset,
ripple strength, and transition slope
from each side of the fade, together
with width and ripple asymmetries.
The 22-dimensional vector extends this
representation with six central-fade
and two boundary descriptors.
All six feature sets use the same
logistic regression classifier.
A radial-basis-function support vector
machine (RBF-SVM) is also evaluated
using the 22-feature representation.

Classification performance is evaluated
using leave-one-participant-out (LOPO)
cross-validation. In each fold, models
are trained using measurements from
two participants and tested on the
remaining participant. This ensures
that measurements from the same
participant do not appear in both
training and testing sets.  Separate models are evaluated for
perpendicular, diagonal, and combined
walking geometries. Geometry-specific
models use only their respective
datasets. For the combined evaluation,
models are trained on both geometries
from the training participants and
tested on both geometries from the
held-out participant. Thus, LOPO
evaluates generalization to unseen
participants within the measured
geometries, rather than to an
unseen walking geometry.

Classification accuracy is computed
from the held-out predictions.
The walking-direction confusion matrices
use the two-feature logistic regression
model, while the remaining feature
sets provide comparative results.
Given the three participants and
exploratory feature comparisons,
the results demonstrate proof-of-concept
performance under the measured
conditions rather than independent
validation of feature selection.

\section{Experimental Results}
\label{sec:results}

\begin{figure}[!t]
\centering
\includegraphics[width=.708\columnwidth]
{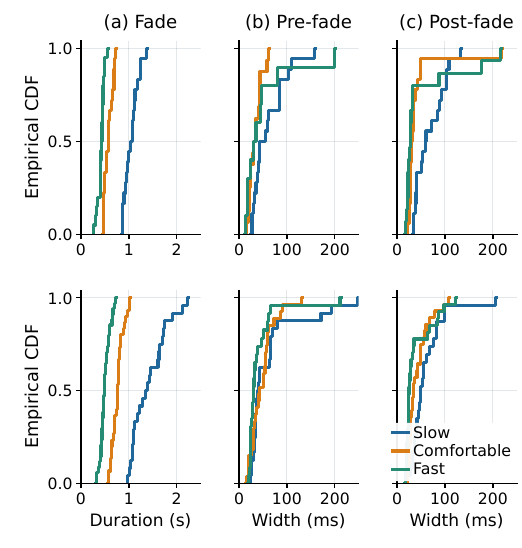}
\caption{Empirical cumulative distributions
of (a) central fade duration, (b) pre-fade
peak half-prominence width, and (c) post-fade
peak half-prominence width for perpendicular
(top, $N=56$) and diagonal (bottom, $N=88$)
DD crossings.}
\label{fig:feature_ecdf}
\end{figure}

We evaluate walking-pace and walking-direction
classification using 144 wall-reflected
60.8-GHz DD crossings from three participants.
Results are reported for perpendicular
($N=56$), diagonal ($N=88$), and combined
($N=144$) datasets under LOPO evaluation.  Fig.~\ref{fig:feature_ecdf} shows the
empirical cumulative distributions of
central fade duration and exterior peak
widths. Slow walking generally produces
longer fades than fast walking, with
comfortable walking between the two
categories. Diagonal crossings also
exhibit longer fades than perpendicular
crossings, indicating dependence on
walking geometry. In contrast, exterior
peak widths exhibit substantial overlap
among pace categories, including slow
and fast walking. This overlap reflects
their dependence on beam-weighted
scattering and interference, rather
than walking pace alone. The stronger
separation in fade duration motivates
its selection as the primary feature
for walking-pace classification.

Table~\ref{tab:classification} summarizes
the LOPO classification results.
Using central fade duration $W_{60}$
alone, binary fast-versus-nonfast
classification achieves 85.7\% (48/56)
and 87.5\% (77/88) accuracy for
perpendicular and diagonal crossings,
respectively. The combined model
achieves 86.1\% (124/144) accuracy.  These results are consistent
with the fade-duration distributions
in Fig.~\ref{fig:feature_ecdf}, where
fast walking generally produces shorter
fades than slow and comfortable walking.
The comparable accuracies across both
geometries suggest that fade duration
provides useful discrimination between
fast and nonfast walking despite
differences in crossing trajectories.  Three-class walking-pace classification
achieves 79.2\% (114/144) accuracy
on the combined dataset. Accuracy
is higher for perpendicular crossings
(82.1\%) than for diagonal crossings
(73.9\%). Although the fade-duration
distributions exhibit clear separation,
distinguishing all three pace categories
is more challenging under LOPO
evaluation. The results also indicate
that classification performance depends
on walking geometry.

\begin{table}[!t]
\centering
\caption{LOPO classification accuracy
for perpendicular, diagonal, and combined
DD crossings.}
\label{tab:classification}
\footnotesize
\setlength{\tabcolsep}{3pt}
\begin{tabular}{lccc}
\toprule
\textbf{Dataset} &
\textbf{Binary} &
\textbf{3-class} &
\textbf{Direction} \\
\midrule
Perp. ($N=56$)
& 85.7\% & 82.1\% & 80.4\% \\
Diag. ($N=88$)
& 87.5\% & 73.9\% & 76.1\% \\
Combined ($N=144$)
& \textbf{86.1\%} & 79.2\% & 77.1\% \\
\bottomrule
\end{tabular}
\end{table}

\begin{figure}[!t]
\centering
\includegraphics[width=\columnwidth]
{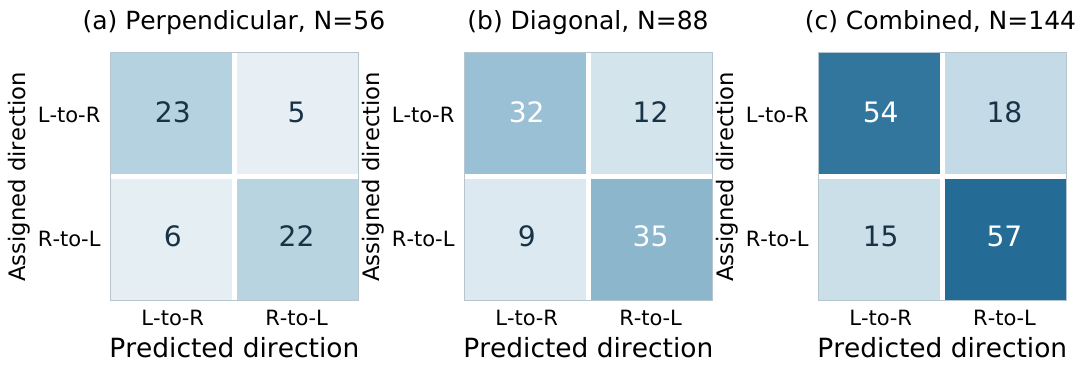}
\caption{Travel-direction confusion matrices
using $A_R$ and $A_S$ under LOPO evaluation
for perpendicular, diagonal
 and combined
DD crossings. Rows indicate reference
directions and columns indicate predictions.}
\label{fig:compact_direction}
\end{figure}

Walking-direction classification exploits
the pre/post-fade ripple-amplitude and
transition-slope asymmetries $A_R$ and $A_S$.
As shown in Table~\ref{tab:classification},
the classifier achieves 80.4\% accuracy
for perpendicular crossings, 76.1\%
for diagonal crossings, and 77.1\%
(111/144) for the combined dataset.
Unlike walking-pace classification,
which relies on central fade duration,
walking-direction classification uses
asymmetries in the interference-related
power fluctuations surrounding the fade.  Fig.~\ref{fig:compact_direction} shows
the corresponding walking-direction
confusion matrices. For the combined
dataset, 54 of 72 left-to-right crossings
and 57 of 72 right-to-left crossings
are correctly classified. The comparable
recognition rates indicate that the
classifier does not strongly favor
either direction. The slightly lower
accuracy for diagonal crossings
indicates some sensitivity to walking
geometry. Overall, the results
demonstrate that beam-shaped
received-power signatures contain
useful walking-direction information
without channel estimation or
Doppler processing.

\begin{figure}[!t]
\centering
\includegraphics[width=.630705\columnwidth]
{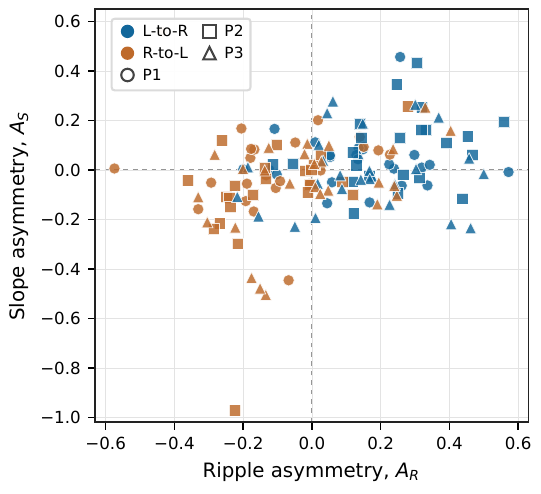}
\caption{Ripple-amplitude asymmetry $A_R$
and transition-slope asymmetry $A_S$
for 144 DD crossings. Both walking
geometries are included.}
\label{fig:direction_feature_space}
\end{figure}

Fig.~\ref{fig:direction_feature_space}
illustrates the distribution of the
two direction features across all
144 crossings. Ripple-amplitude asymmetry
exhibits a noticeable shift between
travel directions, whereas transition-slope
asymmetry shows greater overlap.
Participant-dependent variations are
also evident. These distributions
indicate that the beam-shaped power
signatures contain direction-related
information, although the classes
are not completely separable in the
two-dimensional feature space.
Moreover, the power measurements
do not independently resolve the
underlying scattering paths.

Table~\ref{tab:compact_direction}
compares walking-direction classification
using different feature representations.
Ripple-amplitude asymmetry $A_R$ alone
achieves 75.7\% combined accuracy,
which increases to 77.1\% with the
addition of slope asymmetry $A_S$.
The three- and five-feature models
do not improve this result, while
the 14- and 22-dimensional morphology
representations achieve 72.2\%
and 75.7\%, respectively.  The 22-feature RBF-SVM achieves
79.2\% (114/144) accuracy, compared
with 77.1\% (111/144) for the
two-feature logistic regression model.
This improvement corresponds to only
three additional correct predictions
but requires a larger feature
representation and nonlinear
classification. The two-feature
model therefore provides a compact,
physically interpretable alternative
with comparable performance.
Given the limited number of participants,
these feature comparisons remain
exploratory.

Overall, the results demonstrate
that motion-induced power signatures
from fixed communication beams
contain information about both
walking pace and walking direction.
Both tasks can be addressed using
low-dimensional received-power
features without channel estimation
or Doppler processing.

\section{Conclusion}
We demonstrated device-free human-mobility
sensing using received-power signatures
in a wall-reflected 60.8-GHz mmWave
communication link.  Using 144 crossings from three participants,
leave-one-participant-out evaluation achieved
86.1\% binary and 79.2\% three-class
walking-pace accuracy. Walking-direction
classification achieved 77.1\% accuracy
using two pre/post-fade asymmetry features. These results establish the feasibility
of lightweight power-based sensing for
integrated sensing and communications
without complex channel estimation,
multipath-component resolution, or
Doppler processing.  Future work will
investigate additional randomized trials
and sensing-oriented communication
codebooks with data-bearing signals.

\begin{table}[!t]
\centering
\caption{LOPO walking-direction classification
using different waveform feature sets.}
\label{tab:compact_direction}
\footnotesize
\setlength{\tabcolsep}{3pt}
\begin{tabular}{lccc}
\toprule
\textbf{Feature set} &
\textbf{Perp.} &
\textbf{Diag.} &
\textbf{Combined} \\
\midrule
Ripple asymmetry (1)
& 78.6\% & 75.0\% & 75.7\% \\
Ripple + slope (2)
& \textbf{80.4\%}
& \textbf{76.1\%}
& \textbf{77.1\%} \\
Add minimum timing (3)
& 80.4\% & 76.1\% & 77.1\% \\
Compact features (5)
& 78.6\% & 72.7\% & 77.1\% \\
Exterior morphology (14)
& 66.1\% & 73.9\% & 72.2\% \\
Extended morphology (22)
& 75.0\% & 73.9\% & 75.7\% \\
\bottomrule
\end{tabular}
\end{table}

\section*{Acknowledgment}
This work was supported in part by the
National Science Foundation under Grant
No. 2243089 and the Sacramento State
Research and Creative Activity Faculty
Awards Program.

\balance

\end{document}